\documentstyle[preprint,revtex]{aps}
\begin{document}
\draft
\preprint{October 27, 1994}
\begin{title}
On the Chiral Hubbard Model and the Chiral Kondo Lattice Model
\end{title}
\author{D. F. Wang and C. Gruber}
\begin{instit}
Institut de Physique Th\'eorique\\
\'Ecole Polytechnique F\'ed\'erale de Lausanne\\
PHB-Ecublens, CH-1015 Lausanne-Switzerland.
\end{instit}
\begin{abstract}
In this work, the integrability of the one dimensional chiral Hubbard model
is discussed in the limit of strong interaction $U=\infty$.
The system is shown to be integrable in the sense of
existence of infinite number of constants of motion.
The system is related to a chiral Kondo lattice model
at strong interaction $J=+\infty$.

\end{abstract}

\narrowtext

Hubbard model has been of considerable interests due to its
possible relevance to the high temperature superconductivity.
The one dimensional Hubbard model with nearest neighboring hopping
was solved by Lieb and Wu in 1968 with the help of
Bethe-ansatz\cite{lieb,yang,korepin}. It was shown that the
system exhibits a metal-insulator
phase transition at half-filling even for arbitrarily small
interaction. Away from half-filling, the low lying excitations
of the system have been classified as Luttinger liquid like in
the sense of Haldane.
About two years ago, a one dimensional
$SU(2)$ Hubbard model with only
relativistic right movers was introduced, which reduces,
at half-filling and large but finite on-site energy,
to the $SU(2)$ Haldane-Shastry spin system with $1/r^2$ exchange
interaction\cite{gebhard}. Using
finite size diagonalization result and the information
provided by some special cases, an effective Hamiltonian
was proposed, which was used to provide the full
energy spectrum and the thermodynamics for any on-site energy.
With the help of the effective Hamiltonian,
it was found that at $T=0$ and half-filling, there exists
a critical value $U_c$ at which metal-insulator phase transition
occurs in the system\cite{gebhard}.
It was conjectured that the system is completely integrable
for any on-site energy. However,
a proof for the conjectured energy spectrum and
the thermodynamics, as well as the structure
of the wavefunctions at finite $U$, is still unknown.
The integrability of this Hubbard model remains
an open problem.

In the strong interaction limit $U=\infty$, it was
discovered that the Gutzwiller-Jastrow product wavefunctions
are eigenstates of the chiral Hubbard model, both in the
$SU(2)$ case and in the $SU(N)$ case\cite{wang1}.
In fact all eigenstates
can be expressed in terms of more generalized Jastrow product
wavefunctions.
Furthermore, the $SU(2)$ energy spectrum\cite{wang1} is
the same as conjectured by Gebhard and Ruckenstein
in their original work\cite{gebhard}.
In this work, we would like to discuss
the integrability of the system in the limit
where the interaction between the electrons is infinitely strong
$U=\infty$. Using simple argument, we shall exhibit
an infinite number of constants of motion,
showing that the system is
integrable.

In the second part of our paper, we consider a one dimensional chiral
Kondo lattice. The conduction band has only right moving electrons,
and the electrons interact with each localized impurity moment
through exchange interaction. We identify the chiral Kondo lattice
at $J=+\infty$ with the Hubbard model which has been previously studied.
With this identification, the full energy spectrum, the wavefunctions,
the thermodynamics and the integrability of the system can be obtained
for the Kondo lattice in this limit. In particular, various correlation
functions between the electrons and the impurity spins can be computed
exactly for this system.

The chiral Hubbard model
Hamiltonian is defined on a one dimensional lattice of length $L$:
\begin{equation}
H=\sum_{1\le i\ne j\le L }\sum_{\sigma=1}^N (t_{ij} c_{i\sigma}^\dagger
c_{j\sigma} ) + U \sum_{1\le i \le L} \sum_{\sigma \ne
\sigma'} n_{i\sigma}n_{i\sigma'},
\end{equation}
where the hopping matrix element is given by
$t_{mn} =(-it) (-1)^{(m-n)}[(L/\pi)\sin(\pi(m-n)/L)]^{-1}$.
For this $SU(N)$ system, the spin of the electrons can take
values from $1$ to $N$. In the case of $SU(2)$, it is the Hamiltonian
introduced by Gebhard and Ruckenstein\cite{gebhard}.
In the strong interaction limit $U=\infty$,
at each site there is at most one electron, the number of holes
$N_h$ and the number of the electrons on the lattice $N_e$ are
conserved quantities.
To rewrite the Hamiltonian in more convenient form,
we perform the following unitary transformation $T$:
\begin{eqnarray}
&&c_{x\sigma}^\dagger \rightarrow (-1)^x
e^{-\pi i x/L} c_{x\sigma}^\dagger,\nonumber\\
&&c_{x\sigma}\rightarrow (-1)^x e^{\pi i x/L} c_{x\sigma},
\end{eqnarray}
under which the original Hamiltonian becomes
\begin{equation}
H\rightarrow \bar H=
2t P_G[\sum_{1\le i\ne j\le L} \sum_{\sigma} {z_j \over (z_i-z_j)}
c_{i\sigma}^\dagger c_{j\sigma} ] P_G {\pi \over L} ,
\end{equation}
with $P_G$, the Gutzwiller projector, making sure that there are no
double or multiple occupancies, while $z_x=e^{2\pi i x /L}$, with
$x=1, 2, \cdots, L$.

In the Hilbert space of no double or multiple occupancies,
the electron fields can be rewritten with the
superalgebra representations,
\begin{eqnarray}
P_G(i) c_{i\sigma}^\dagger P_G(i) = f_{i\sigma}^\dagger b_i,\nonumber\\
P_G(i)c_{i\sigma}P_G(i) = b_i^\dagger f_{i\sigma},
\end{eqnarray}
where the $f$ fields are fermions, the $b$ fields are bosons,
with the constrain
$\sum_{\sigma} f_{i\sigma}^\dagger f_{i\sigma} + b_i^\dagger b_i =1$.
$P_G(i)$ is the Gutzwiller projector operator on the site
$i$ for the electron operators $c$ and $c^\dagger$.
Any state vector of the Hilbert space can be written as
\begin{equation}
|\phi>=\sum_{\{x\sigma\},\{y\}} \phi (\{x\sigma\}, \{y\}) \prod_{i=1}^{N_e}
f_{x_i\sigma_i}^\dagger \prod_{j=1}^{N_h} b_{y_j}^\dagger|0>,
\end{equation}
where the amplitude $\phi$ is symmetric in the coordinates
$\{y\}$ of the $b$ bosons, while antisymmetric when exchanging
the spin and positions $x_i\sigma_i, x_j\sigma_j$ of two $f$ fermions.
Here, $(q_1, q_2,\cdots,q_L) = (x_1,\cdots,x_{N_e},y_1,\cdots,y_{N_e})$
span the full chain.
The eigenenergy equation of the system can then
be written in the first quantized
form as follows:
\begin{equation}
 {2t \pi \over L}[\sum_{i=1}^{N_e} \sum_{j=1 (\ne i)}^L {Z_j\over (Z_i-Z_j)}
M_{ij}
-{1\over 2} \sum_{1\le i\ne j\le N_e} P_{ij}^\sigma ]  \phi (\{q\},\{\sigma\})
=E \phi(\{q\},\{\sigma\}),
\label{eq:eigen}
\end{equation}
where $Z_i=e^{2\pi i q_i/L}$,
the operator $M_{ij}$ exchanges the position variables
$q_i$ and $q_j$, the operator $P_{ij}^\sigma$ exchanges
the $f$ fermion spin variables $\sigma_i$ and $\sigma_j$.
These two operators commute with each other, as they act on
different groups of variables of the wavefunction.

Following the ideas of Ref.\cite{fowler},
we define generalized momentum operators $\Pi_i$, with $i=1, 2, \cdots, N_e$,
\begin{equation}
\Pi_i=\sum_{j=1(\ne i)}^L V_{ij} M_{ij},
\label{eq:mom}
\end{equation}
with $V_{ij}= Z_j/(Z_i-Z_j)$. With this special form of $V_{ij}$,
the generalized momentum operators satisfy the commutation relation
\begin{equation}
[\Pi_i,\Pi_j]=M_{ij} \Pi_i -\Pi_i M_{ij},
\label{eq:com}
\end{equation}
where $1\le i\ne j\le N_e$.
We then introduce the following hermitian operators
\begin{equation}
A_n=\sum_{s=1}^{N_e} \Pi_s^n,
\end{equation}
where $n=0, 1, 2, \cdots, \infty$, and the sum $s$ is over the
electrons, i.e., from $1$ to $N_e$.
In particular, the Hamiltonian is given by
\begin{equation}
\hat H = {2\pi t\over L}
[A_1-{1\over 2} \sum_{1\le i\ne j\le N_e} P_{ij}^\sigma].
\label{eq:hamil}
\end{equation}
Using the commutation relations Eq.~(\ref{eq:com}) it can be shown
that all the operators $A_n's$ commute with each other.
Furthermore, the action of $A_n$ on some
amplitude $\phi$ does not change the symmetry properties,
i.e., the resultant wavefunctions
remains symmetric in the $b$ boson positions, and antisymmetric
when exchanging any pair of the $f$ fermion positions and spins
simultaneously. It is straightforward to prove the following relations
\begin{eqnarray}
M_{ij} [A_n\phi] &&=[A_n\phi], \, N_e+1\le i\ne j \le L\nonumber\\
M_{ij}P_{ij}^\sigma [A_n \phi]&& = (-1) [A_n\phi], \, 1\le i\ne j\le N_e.
\end{eqnarray}
Since all the operators $A_n's$
commute with the Hamiltonian given by the
Eq.~(\ref{eq:hamil}), we thus have an infinite set of
conserved physical quantities of the system, showing that the system
is indeed completely integrable.
With these $A_n's$, we can construct corresponding quantities
written in second quantized language, which commute among themselves
and with the Hamiltonian $\bar H$. Carrying out the unitary
transformation $T^{-1}$, it is straightforward to convert them, so that
the resultant quantities are constants of the original Hamiltonian $H$.

One interesting observation is that the mutually commuting
hermitian quantities $A_n$'s are also
the invariants of the long range supersymmetric t-J model\cite{kur},
i.e., $[A_n, H_{t-J}]=0$. As noted previously,
the physical quantities,
$I_n=\sum_{i=1}^L \Pi_i^n$,
with $n=0, 1, 2, \cdots$,
commute with each
other, and with the supersymmetric t-J model Hamiltonian.
The two families of quantities $\{I_n\}, \{A_n\}$ are independent
of each other, in the sense that we can not write any member of
one group in terms of a linear combination of members of the other
group. Furthermore, by explicit computation, one can show that
they do not commute, e.g., $[A_n, I_m]\ne 0$.
Therefore, it is clear that the previous family
of conserved quantities $\{I_n\}$, although providing
a proof of the integrability of the t-J model,
does not exhaust all the constants of motion.
The two infinite symmetries of the system do not commute
with each other\cite{lieb2}.
It would be very interesting to find larger group of
mutually commuting constants of motion.
This is actually one of the most fundamental questions
encountered in study of the quantum integrable systems.
That is, when one has found an infinite number of simultaneous
constants of motion,showing that the system is integrable,
it is still not sure
that the infinite set would contain $\it all$ possible
simultaneous constants of motion of the system.

In the following, we will show how the results of the chiral Hubbard
model may be generalized to a one dimensional Kondo lattice model.
The Kondo lattice model has been an interesting model for the study
of heavy fermion systems\cite{lee}. In this model, the system has
an array of localized impurity moments, the conduction electrons
interact with the local moments through spin exchange. In general,
the conduction band is best described by the tight banding picture,
in which we have both right and left movers. However,
in the following, we assume that the electrons
propagate in only one direction. This chiral
Kondo lattice model is defined on a one dimensional lattice,
with the Hamiltonian:
\begin{equation}
H=\sum_{k}\sum_{ \sigma=\uparrow,\downarrow}
e(k) c_{k\sigma}^\dagger c_{k\sigma} + J \sum_{i=1}^L
c_{i\alpha}^\dagger {\vec \sigma_{\alpha\beta} \over 2} c_{i\beta}
\cdot {\vec S_f(i)},
\end{equation}
where the conduction band spectrum is $e(k)=-tk$, with
$k={2\pi K\over L}$, $-(L-1)/2\le K\le (L-1)/2$,
in the momentum space. $J$ is the coupling
constant between the local impurity moments and the conducting electrons.
The local moments are described by the spin 1/2 operators, that is,
$[S_f^x(k), S_f^y(k)]=iS_f^z(k)$ (plus two other commutation
relations obtained by the cyclic permutations of $x,y,z$),
with the relation $\vec S_f^2(k) =3/4$, for all the sites $k=1, 2, \cdots, L$.
The Hamiltonian may also be written in the following way:
\begin{equation}
H=\sum_{1\le i \ne j \le L}\sum_{ \sigma=\uparrow,\downarrow}
t_{ij} c_{i\sigma}^\dagger c_{j\sigma} + J \sum_{i=1}^L
c_{i\alpha}^\dagger {\vec \sigma_{\alpha\beta} \over 2} c_{i\beta}
\cdot {\vec S_f(i)},
\end{equation}
where $t_{mn}=(-it) (-1)^{(m-n)}[(L/\pi)\sin(\pi(m-n)/L)]^{-1}$.

When the interaction of the electron and the impurity is very strong,
i.e. $J=+\infty$, we can map the system onto the above chiral Hubbard model
with infinity repulsion.
Indeed, when there are $N_e$ electrons
on the lattice $L$, with $N_e\le L$, then each electron will attempt to form
a singlet with the impurity spin at each site, to lower the energy
of the system as much as possible, and some unpaired
impurity spins are left over on the lattice.
The Hilbert space at each site
can be either a unpaired impurity spin or a singlet of electron-impurity
bound state. Due to the hopping of the conduction electrons, the singlets
can hop on the lattice.
In this case, the basis vectors can be written as
\begin{eqnarray}
|\alpha>=2^{-N_e/2} \left[ \prod_{i=1}^{N_e} (1-P_{\gamma_i\beta_i}) \right]
&&c_{x_1\gamma_1}^\dagger c_{x_2\gamma_2}^\dagger \cdots
c_{x_{N_e} \gamma_{N_e}}^\dagger |0>\nonumber\\
&&\bigotimes
|\sigma_1, \sigma_2, \cdots, \beta_1, \cdots, \beta_2, \cdots, \sigma_{L-N_e}>,
\end{eqnarray}
where the singlets are located at positions $\{x\}=(x_1<x_2<\cdots<x_{N_e})$,
the unpaired impurity spins $(\sigma_1, \sigma_2, \cdots, \sigma_{L-N_e})$
are positioned at sites $\{y\}=(y_1< y_2<\cdots<y_{L-N_e})$.
Here, the operator $P_{\gamma_i\beta_i}$ permutes the spin indices
$\gamma_i$ and $\beta_i$, to form a singlet of electron and impurity at site
$x_i$.
With $P$ the projector onto this subspace,
the Hamiltonian takes the form:
\begin{equation}
\tilde H = P H P = P T P + c,
\end{equation}
where $T$ is the kinetic energy of the conduction electrons,
the infinite constant $c=(-J/4)N_e$ only shifts the origin of
the energy of the system, a reference energy
which is unimportant physically.
In the space where the z-component of the total spin is fixed,
that is, $S_z=M$, the number of the unpaired up-spin impurities
is $A=M+(L-N_e)/2$, the number of the unpaired down-spin impurities
is $B=-M +(L-N_e)/2$.
$C_L^{N_e} \times C_{L-N_e}^A$ is the
size of the Hilbert space. Any eigenstate of the
Hamiltonian $H_1= PTP$ can be written as a linear combination
of the basis vectors,
\begin{equation}
|\phi> =\sum_{\alpha} C(\alpha) |\alpha>.
\end{equation}
We can identify the singlets as spinless fermions,
the unpaired impurities as hard core spin 1/2 bosons
hopping on the lattice. Let us consider a system described
by the following Hamiltonian:
\begin{equation}
h=(1/2) \sum_{i\ne j, \sigma} P_g (t_{ij} b_{j\sigma}^\dagger b_{i\sigma}
g_i^\dagger g_j) P_g
\end{equation}
where the $b$ fields are bosonic, $g$ fields are fermionic,
$b$ fields commute with the $g$ fields,
and the Gutzwiller projector
$P_g=\prod_{i=1}^L[\delta_{1,g_i^\dagger g_i
+\sum_{\sigma=\uparrow,\downarrow}
b_{i\sigma}^\dagger b_{i\sigma}}] $.
The basis vectors may be represented as follows:
\begin{equation}
|\bar \alpha> = g_{x_1}^\dagger g_{x_2}^\dagger \cdots g_{x_{N_e}}^\dagger
b_{y_1\sigma_1}^\dagger b_{y_2\sigma_2}^\dagger
\cdots b_{y_{L-N_e} \sigma_{L-N_e}}^\dagger |0>.
\end{equation}
One can show that the systems described by $H_1$ and $h$ are isomorphic
to each other, by verifying the following matrix elements:
\begin{equation}
<\beta|H_1|\alpha> = <\bar \beta| h | \bar \alpha>,
\end{equation}
where there is the one-to-one correspondence
$|\alpha> \leftrightarrow |\bar \alpha>$ for the basis vectors.
The Hamiltonian $h$ is equivalent to the following Hamiltonian:
\begin{equation}
h=(1/2) \sum_{i\ne j, \sigma}
(-t_{ji}) P_F F_{i\sigma}^\dagger F_{j\sigma} P_F,
\end{equation}
where $P_F = \prod_{i=1}^L P_F(i)$,
and $P_F(i) = (1-F_{i\uparrow}^\dagger F_{i\uparrow}
F_{i\downarrow}^\dagger F_{i\downarrow})$, and $L-N_e$ is the number
of the $F$ fermions on the lattice.

With the above identification, we have mapped the chiral Kondo lattice model
onto the chiral Hubbard model with strong repulsion. Therefore, an
infinite number of mutually commuting invariants can be obtained
for the Kondo lattice model. The wavefunctions and the thermodynamics of
the system may be read from previous results\cite{wang1}.
Any state vectors can be written as
\begin{equation}
|\phi> =\sum_{\{X\}, \{Y\}} \Phi (\{X\}, \{Y\} )
\prod_{i=1}^{\bar Q} F_{Y_i\downarrow} \prod_{j=1}^B
F_{X_j\downarrow}^\dagger F_{X_j\uparrow} |P>,
\end{equation}
where $|P> =
\prod_{i=1}^L F_{i\uparrow}^\dagger |0>$, $\bar Q = N_e$
is the number of $g$ fermions,
$B=-M+(L-N_e)/2$ is the number of down-spin $b$ bosons,
the amplitude $\Phi$ is antisymmetric
in the positions $\{Y\}$, while symmetric in the positions
$\{X\}$. The following Jastrow wavefunctions are eigen
states of the Hamiltonian:
\begin{equation}
\Phi (\{X\}, \{Y\}) = e^{(2\pi i /L) (m_s \sum_{i} X_i + m_h \sum_{j} Y_j )}
\prod_{i<j} d^2(X_i-X_j) \prod_{i<j} d(Y_i-Y_j) \prod_{i, j} d(X_i-Y_j),
\end{equation}
with $d(n)=\sin(\pi n/L)$. The quantum numbers
$m_s, m_h$ are integers or half-integers, which make sure of the periodic
boundary conditions, satisfying the following constrains
\begin{eqnarray}
&&|m_h| \le L/2 -(B+\bar Q)/2,\nonumber\\
&&|m_h-m_s-L/2 | \le L/2 -(A
+\bar Q)/2,
\end{eqnarray}
with the eigen-energies given by
\begin{equation}
E(m_s,m_h)=-(2\pi  t/L) [2 m_h -m_s +L/2] \bar Q (1/2).
\end{equation}
The full spectrum of the system takes the following form
\begin{equation}
E =-(2 \pi t /L) [  \sum_{i=1}^{\bar Q} n_i + \sum_{\mu = 1}^{\bar Q}
m_{\mu}] (1/2).
\end{equation}
Here, the integers (or half integers ) satisfy the conditions
$|n_i| \le L/2 -(A +\bar Q)/2, |m_{\mu} |
\le L/2 -(B+\bar Q)/2$, where $n_i \le n_{i+1}$ and $m_{\mu} \le m_{\mu +1}$.
This result shows that the spectrum is invariant
when changing the sign of $t$.

The Jastrow product wavefunctions of the unpaired impurity spins
and the singlets are typical RVB-type wavefunctions. Various
correlation functions of the impurity spins and the singlets
can be computed $\it exactly$, by trivially generalizing
Forrester's work to this case. It should be remarked
that the far away unpaired impurity spins are
also strongly correlated with each other,
because of the fact that only right movers exist in the conduction band.
At half-filling, the system is obviously an insulator, since each electron
forms a singlet at each site and the singlets can not hop from
one site to another.

For this chiral Kondo lattice model, the conduction electrons
move only in one direction. We can anticipate many physical properties for
the system even at finite coupling constant $J$. Away from half-filling,
one would expect the system to be in a metallic state.
At half-filling and for sufficiently large $J$, the system is expected
to be insulating. In this model,
the chirality of the conduction band won't prevent
the system from becoming insulating, unlike in some other
situations, such as in the edges of the Fractional Quantum Hall Effect,
where the chiral Luttinger liquid
won't become localized under any randomness,
due to lack of backscattering of the quasiparticle\cite{wen}.
In our case, although the electrons are moving in only one direction,
the mechanism for localization is very different. The electron
always feels the exchange interaction of the
impurity spin, through the spin exchange interaction.
For $J$ large enough,
each electron will attempt to form a localized singlet with
each impurity spin, therefore, at half-filling,
to transfer one electron from one site
to another would break two singlets, causing
a charge gap of order of $O(J)$, and the system
would be in an insulating state.
At $J=0$, the system is a simple Fermi liquid. One would thus
expect that there exists a critical coupling $J_c$, where
the system exhibits metal-insulator phase transition at half-filling.
An interesting question is whether
any infinitesimal small $J$ would drive the conducting band to an insulating
state at half-filling, i.e. $J_c = 0^+$.
If $J_c\ne 0^+$, one would expect $J_c\sim |t|$ by dimensional
analysis, and the system is metallic for $0<J<J_c$, while
it becomes insulating for $J_c<J$. Metal-insulator phase
transition would also occur at half-filling when changing
$|J|$ for ferromagnetic interaction between the conduction electrons
and the impurity spins. Further work is necessary to locate the
critical coupling $J_c$.
It might also be interesting to see
whether the model at finite $J$ belongs to
Jastrow-integrable type.

In summary, we have obtained an infinite number of constants of motion
for the one dimensional chiral Hubbard model in the strong interaction
limit $U=\infty$.
We have also shown that this model is equivalent to
the one dimensional chiral Kondo lattice
model at $J=+\infty$.
It seems that the integrability condition might
be investigated for finite on-site energy, using a similar approach.
However, we have not succeeded in doing so for finite on-site energy.
It has also seemed that the finite $J$ chiral Kondo lattice model
is very probably to be integrable, as the conduction electrons
move in one direction, only exchanging spins with the local moments.
In the continuum limit, loosely speaking, the many particle
scattering matrices are very probable to be factorized into two
body ones with a continuum relativistic electrons, and the situation
of scattering matrices of the electrons off one impurity might be
similar to the S-matrices of the solution of Andrei and $\it al$\cite{andrei}.
Further work, either numerically or analytically,
is necessary for its solvability evidence.

We wish to thank Prof. Ph.Choquard for stimulating
conversations about constants of motion of classical
and quantum Calogero-Sutherland type models, and about
their possible relevance
to practically measurable quantities in real condensed
matter systems. It is also a great pleasure to thank Prof. E. H. Lieb
for interesting discussions.
The financial support from the Swiss National Foundation for Science
is gratefully acknowledged.


\begin{references}
\bibitem{lieb2} Similar situation might happen in conventional
one dimensional Hubbard model of nearest-neighbor hopping and
finite on-site energy $U$, E. H. Lieb (private conversation).
\end{references}
\end{document}